\documentclass[a4paper,cite,11pt,psnfss]{article}

\usepackage[latin1]{inputenc}
\usepackage{comment}
\usepackage{ulem}
\usepackage{adjustbox}
\usepackage{appendix}
\usepackage{framed,epsfig,pgfplots,amsmath}
\usepackage{tikz}
\usepackage{tikz-feynhand}
\usepackage{autobreak}
\usepackage{tikz-feynman}
\usetikzlibrary{snakes,arrows,shapes,positioning,automata,backgrounds,calc,er,patterns}

\usepackage{mathtools, amsmath, amsthm, amssymb,amsfonts}
\usepackage{makecell}
\usepackage{bm}

\usepackage[usenames,dvipsnames,tree]{pstricks}

\usepackage{graphicx,color,pst-plot}

\usepackage{latexsym,amsmath,amsfonts,amssymb,amstext,mathrsfs,upgreek}

\newcommand{\hyper}[5]{\;_{#1}{\rm F}_{#2} \left(\left.\begin{} {#3}
\\ {#4} \end{matrix}\right| #5\right) }

\newmuskip\pFqmuskip
\newcommand*\pFq[7][8]{%
  \begingroup 
  \pFqmuskip=#1mu\relax
  \mathchardef\normalcomma=\mathcode`,
  \mathcode`\,=\string"8000
  \begingroup\lccode`\~=`\,
  \lowercase{\endgroup\let~}\pFqcomma
  {}_{#2}{#3}_{#4}{\left[\genfrac..{0pt}{}{#5}{#6};#7\right]}%
  \endgroup
}

\newcommand{\pFqcomma}{{\normalcomma}\mskip\pFqmuskip}

\allowdisplaybreaks[1]

\begin{document}

\begin{titlepage}
\begin{flushright}
%
\end{flushright}
\begin{flushright}
\end{flushright}

\vfill

\begin{center}

{\Large\bf Multiple Mellin-Barnes integrals with straight contours \\}
 \vspace{0.5cm}
{\Large\bf }

\vfill

{\bf Sumit Banik$^{a\dagger}$ and Samuel Friot$^{b,c\ddagger}$}\\[1cm]
{$^a$ Centre for High Energy Physics, Indian Institute of Science, \\
Bangalore-560012, Karnataka, India}\\[0.5cm]
{$^b$ Universit\'e Paris-Saclay, CNRS/IN2P3, IJCLab, 91405 Orsay, France } \\[0.5cm]
{$^c$ Univ Lyon, Univ Claude Bernard Lyon 1, CNRS/IN2P3, \\
 IP2I Lyon, UMR 5822, F-69622, Villeurbanne, France}\\[0.5cm]
\end{center}
\vfill

\begin{abstract}
We show how the conic hull method, recently developed for the analytic and non-iterative evaluation of multifold Mellin-Barnes (MB) integrals, can be extended to the case where these integrals have straight contours of integration parallel to the imaginary axes in the complex planes of the integration variables. MB integrals of this class appear, for instance, when one computes the $\epsilon$-expansion of dimensionally regularized Feynman integrals, as a result of the application of one of the two main strategies (called A and B in the literature) used to resolve the singularities in $\epsilon$ of MB representations. 
We upgrade the \textit{Mathematica} package \texttt{MBConicHulls.wl} which can now be used to obtain multivariable series representations of multifold MB integrals with arbitrary straight contours, providing an efficient tool for the automatic computation of such integrals. This new feature of the package is presented, along with an example of application by calculating the $\epsilon$-expansion of the dimensionally regularized massless one-loop pentagon integral in general kinematics and $D=4-2\epsilon$.
\end{abstract}

\vspace{1cm}

\small{$\dagger$ sumitbanik@iisc.ac.in}

\small{$\ddagger$ samuel.friot@universite-paris-saclay.fr}

\end{titlepage}

\section{Introduction}

Mellin-Barnes (MB) integrals appear in many branches of mathematics and physics: asymptotics \cite{Paris&Kaminsky}, hypergeometric function theory \cite{KdF,Exton,Marichev}, quantum field theory \cite{Smirnov:2012gma}, electromagnetic waves in turbulence \cite{Sasiela}, etc. In all these domains one makes use of these integrals as a powerful computational tool. This has, for instance, been recalled very recently in the context of particle physics in \cite{Belitsky:2022gba,Dubovyk:2022obc}.  Focusing on the latter domain, it is fair to say that the community of high energy physics has been particularly involved in the study of MB integrals and, during the last two decades, in the development of softwares dedicated to their application to the calculations of Feynman integrals (see \cite{Smirnov:2012gma,Dubovyk:2022obc} and references therein).
Indeed in the phenomenology of particle physics, the computational need is so huge and the complexity of the calculations so high that it is often impossible to avoid computers for the calculations. MB integrals have not been an exception to this rule and powerful softwares are now available to ease their use \cite{MBtools}.

Although MB integrals have been widely studied, the problem of their analytic and numerical evaluation is still an active field of research, in particular when these integrals are multifold. In \cite{Ananthanarayan:2020fhl}, an important progress has been achieved in this context, where the first systematic method of computing multifold Mellin-Barnes (MB) integrals analytically, in a non-iterative way, has been presented, along with the powerful and user-friendly \textit{Mathematica} package \texttt{MBConicHulls.wl} allowing applications of this technique in an automatic way. From this approach, one obtains series representations of multiple MB integrals by computing the latter using multidimensional residue theory. In general, these series representations have the form of linear combinations of multivariable hypergeometric series (and/or derivatives of such series with respect to their parameters). These representations are useful in various fields of physics and mathematics, such as in quantum field theory (for the computation of Feynman integrals, as mentioned above) or in the theory of multivariable hypergeometric functions (for the study of their transformation theory). 
One strength of the approach of \cite{Ananthanarayan:2020fhl} is that, in the common situation where several series representations of the object under study, all being convergent\footnote{When all series representations are converging, this is denoted as a degenerate case \cite{Ananthanarayan:2020fhl}.}, coexist, it bypasses one major difficulty met in other calculational approaches such as, in physics, the negative dimension approach \cite{Halliday:1987an,dunne1987negative} (see also \cite{DelDuca:2009ac}), the MB approach developed in \cite{Sasiela}, the method of brackets \cite{Gonzalez:2010uz} or other more recent techniques, such as the one developed in \cite{Loebbert:2019vcj} based on Yangian symmetry. All these methods need a detailed convergence analysis of the complete set of series involved in the calculation, which can be more than thousands in non-trivial cases \cite{Loebbert:2019vcj,Ananthanarayan:2020ncn}. In contrast, in  \cite{Ananthanarayan:2020fhl}  the series representations are not obtained from a convergence analysis, but from a simple geometrical approach based on the study of specific intersections of conic hulls associated with the MB integral. This allows one to derive the series representations in complicated cases with many variables where the other methods above fail. Another important advantage of the approach of \cite{Ananthanarayan:2020fhl}, compared to all other methods, is that, in the case where one is interested in the convergence region of a given series representation, one can focus on a single master series and not on all the series that form the series representation (because from the conic hull approach one master series can be obtained for each of the various series representations). This considerably simplifies the convergence analysis.

The first applications of the computational method of \cite{Ananthanarayan:2020fhl} have been published in \cite{Ananthanarayan:2020ncn,Ananthanarayan:2020xpd} and \cite{Ananthanarayan:2020fhl}, where complicated conformal Feynman integrals have been evaluated analytically for the first time, and in \cite{Friot:2022dme}, where it was shown, on the example of Srivastava's triple hypergeometric function $H_C$, that this method can be a powerful tool for the derivation of linear transformations of multivariable hypergeometric functions. In all these calculations, the contours of integrations of the involved MB integrals are such that they do not ``split'' the sets of poles of each of the gamma functions that belong to the numerator of the MB integrand in subsets, because this is the way the fundamental objects under study, \textit{i.e.} (dimensionally regularized) Feynman integrals and hypergeometric functions, are well-defined in terms of MB integrals. This condition, in general, forces the contours to be non-straight.

In this paper, we consider the different situation where the contours of the multifold MB integrals can be any straight lines parallel to the imaginary axes in the complex planes of the integration variables (these lines, obviously, avoid the poles of the MB integrand). In mathematics, this is a problem of general interest. In quantum field theory, this kind of MB integrals appear when one computes the $\epsilon$-expansion of dimensionally regularized Feynman integrals following the  MB representation approach summarized in Chapter 5 of \cite{Smirnov:2012gma}. Indeed, in this approach, one performs the $\epsilon$-expansion at the MB integral level, which asks to resolve the problem of $\epsilon$-singularities following two main strategies (called A and B in the literature \cite{Smirnov:2012gma,Smirnov:2009up}), both leading to multifold MB integrals with straight contours having the shape described above.

MB integrals with straight contours generally have the sets of poles of some or all the gamma functions of the numerator of their integrand split in subsets by the contours. This prevents from directly applying the method of \cite{Ananthanarayan:2020fhl} to the computation of such integrals: it is first necessary to perform some transformations of the MB integrand, as we show in the next section. Therefore, in order to deal with these cases, we have adapted the \texttt{MBConicHulls.wl} package (which can be downloaded from \cite{git}) by adding an option in the code allowing the user to define straight contours of integration. Once the straight contours are specified by the user, the package performs the corresponding necessary transformations automatically and the results of the computation of the multifold MB integral can then be derived in an automatic way, as done with the original version of the package presented in  \cite{Ananthanarayan:2020fhl}.

We explain these considerations in detail in Section \ref{section1}, on the simple example of a 2-fold MB integral where the calculations of the transformations are done by hand, whereafter we present the corresponding syntax that has to be used when using the new version of our package for an automatic treatment of the same calculations. In Section \ref{section2} we show a non-trivial application of our method by computing the $\epsilon$-expansion of the dimensionally regularized massless one-loop pentagon integral in general kinematics and $D=4-2\epsilon$. This calculation involves several MB integrals, up to 4-fold, with straight contours, and we show how one can easily obtain different series representations of the pentagon from these integrals. To our knowledge, these results have not been previously published in the literature.  An alternative computational approach of multifold MB integrals with straight contours has been developed in \cite{Ochman:2015fho} and automatized in the \texttt{MBsums.m} \textit{Mathematica} package.  It is based on an iterative approach: the MB integrals are evaluated sequentially. In contrast, our \texttt{MBConicHulls.wl} package  computes multifold MB integrals in a non-iterative way. As one will see, the pentagon example of Section \ref{section2} gives us the opportunity to compare these two different approaches and packages. 
The conclusions of our paper then follow.

\section{$N$-fold MB integrals with straight contours\label{section1}}

The general form of the $N$-fold MB integrals with straight contours that we consider in this work is
\begin{align} \label{N_MB}
    I &(x_1,x_2,\cdots ,x_N) = \int\limits_{c_1-i \infty}^{c_1+i \infty} \frac{ \text{d} z_1}{2 \pi i} \cdots \int\limits_{c_N-i \infty}^{c_N+i \infty}\frac{ \text{d} z_N}{2 \pi i}\,\,  x^{z_1}_{1} \cdots x^{z_N}_{N} \frac{\prod\limits_{i=1}^{k} \Gamma^{a_i}({\bf e}_i\cdot{\bf z}+g_i)}{\prod\limits_{j=1}^{l} \Gamma^{b_j}({\bf f}_j\cdot{\bf z}+h_j)}
\end{align}
where $a_i , b_j, k, l$, $N$ are positive integers, ${\bf z}=(z_1, \cdots, z_N)$, ${\bf e}_i$ and ${\bf f}_j$ are $N$-dimensional real vectors while $g_i$, $h_j$ and the variables $x_1 , \cdots , x_N$ can be complex. 
The integration contours are such that $\Re(z_i)=c_i$ for $i=1,...,N$, \textit{i.e.} they form straight lines parallel to the imaginary axes in each of the $z_i$ complex planes. We restrict our discussion in this paper to the case where the vector ${\bf\Delta}\doteq\sum_ia_i{\bf e}_i-\sum_jb_j{\bf f}_j={\bf 0}$. This is the degenerate case that we mentioned in the introduction and which, to our knowledge, includes the class of MB representations appearing in Feynman integral calculus. The non-degenerate case is presently under study.

As mentioned in the introduction, in the original computational approach of multifold MB integrals presented in \cite{Ananthanarayan:2020fhl} (we do not recall this approach here and refer the reader to  \cite{Ananthanarayan:2020fhl} and to \cite{Ananthanarayan:2020xpd} for technical details), it is assumed that the contours do not split, for each of the gamma functions of the numerator of the MB integrand, their set of poles in different subsets. 
An equivalent way to formulate this assumption, in the straight contour case described above, is that the real part of the arguments of each of the gamma functions of the numerator of the MB integrand must be positive for any values of the integration variables running on the contours (this, in passing, is a necessary requirement to derive well-defined MB representations with straight contours for Feynman integrals \cite{Tausk:1999vh,Anastasiou:2005cb}).
Obviously, when computing multiple MB integrals with straight contours as given in Eq.(\ref{N_MB}), this requirement is in general not satisfied, hence one cannot directly apply the method of \cite{Ananthanarayan:2020fhl} for the computation of these integrals. It is first necessary to transform those gamma functions that do not have their arguments with positive real parts, in such a way that,  for the chosen straight contours, they satisfy this condition. This can be done using the generalized Euler reflection formula, as we show here on a simple example.

\begin{figure}[h]
\centering
\includegraphics[width=7cm, height=7cm]{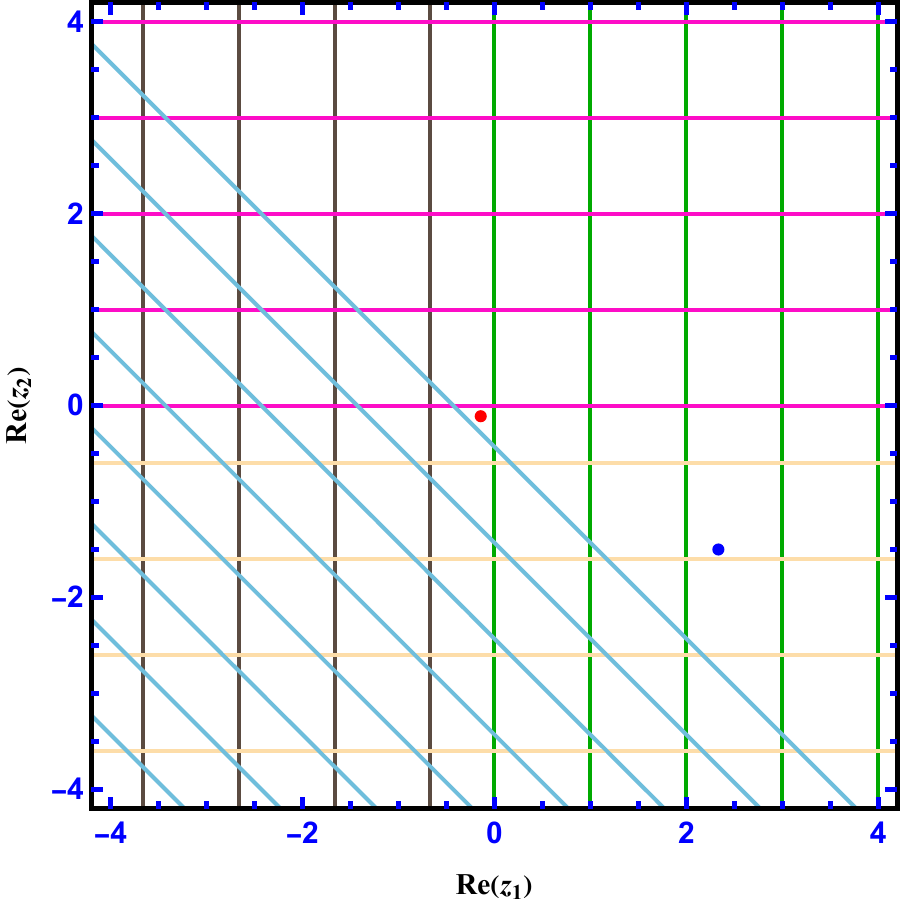}
\hspace{1cm}
\includegraphics[width=7cm, height=7cm]{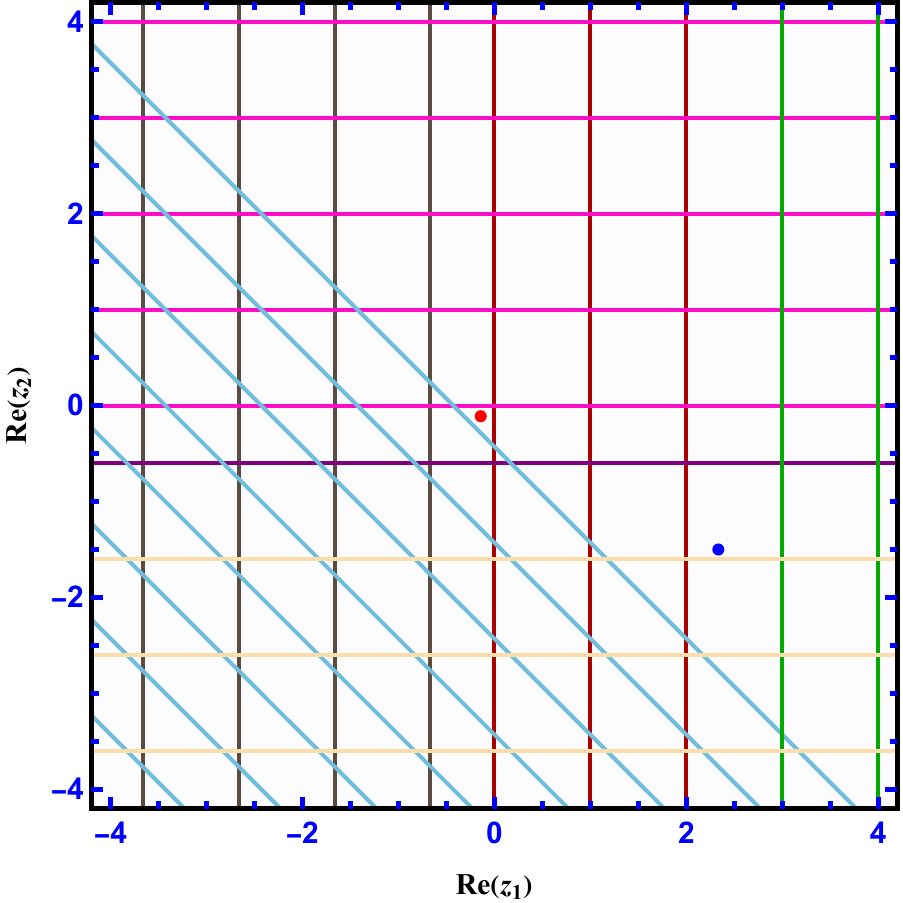}
\caption{Singular structure, in the $(\Re(z_1),\Re(z_2))$-plane, of the integrand of \textit{Left}:  Eq.(\ref{2_MB}) and \textit{Right}: Eq.(\ref{2_MB2}). All the poles (represented as singular lines in the figures) of a given gamma function of the numerator of the corresponding MB integrands are plotted with the same color (for instance, the poles of $ \Gamma(\frac{3}{7}+z_1+z_2)$ are the oblique lines shown in light blue). In the \textit{Right} figure, $\Gamma(-2+z_1)$ (resp. $\Gamma(-\frac{3}{5}-z_2)$) has only 3 (resp. 1) singular lines, as the others are cancelled by the denominator of the MB integrand. The red point is $(c_1,c_2)=(-\frac{1}{7},-\frac{1}{9})$ and the blue one is $(c_1,c_2)=(\frac{7}{3},-\frac{3}{2})$.\label{Singular}}
\end{figure}

Let us consider the following 2-fold MB integral
\begin{align} \label{2_MB}
    I &(x_1,x_2) = \int\limits_{c_1-i \infty}^{c_1+i \infty} \frac{ \text{d} z_1}{2 \pi i} \int\limits_{c_2-i \infty}^{c_2+i \infty}\frac{ \text{d} z_2}{2 \pi i}\,\, (-x_1)^{z_1} (-x_2)^{z_2} \Gamma(-z_1)\Gamma(-z_2) \frac{ \Gamma(\frac{3}{7}+z_1+z_2)\Gamma(\frac{2}{3}+z_1)\Gamma(\frac{3}{5}+z_2)}{\Gamma(\frac{1}{2}+z_1+z_2)} 
\end{align}
If one chooses the contours of integration such that all the five gamma functions in the numerator of the MB integrand satisfy the positivity constraint of the real part of their respective argument for any $z_1$ and $z_2$ running on the contours, for instance by fixing $c_1=-\frac{1}{7}$ and $c_2=-\frac{1}{9}$, then, up to an overall factor, one recognizes the MB representation of the Appell $F_1$ function
\begin{align} \label{2_MB_F1}
    I &(x_1,x_2) = \frac{ \Gamma(\frac{3}{7})\Gamma(\frac{2}{3})\Gamma(\frac{3}{5})}{\Gamma(\frac{1}{2})} F_1\left(\frac{3}{7},\frac{2}{3},\frac{3}{5};\frac{1}{2};x_1,x_2\right)
\end{align}
For this choice of straight contours the sets of poles of each of the gamma functions in the MB integrand are not split in different subsets. One can see this fact by looking at  Fig. \ref{Singular} \textit{Left} where the red point $(c_1,c_2)=(-\frac{1}{7},-\frac{1}{9})$ is surrounded by the singular lines of the five gamma functions of the numerator of the MB integrand in a particular way, whose visualization we have tried to ease by giving an identical color to all the singular lines of a given gamma function. Indeed, it is clear from the picture that the point $(-\frac{1}{7},-\frac{1}{9})$ is not located between two singular lines of the same color. This is what is meant when we say that, for each of the gamma functions in the numerator of the MB integrand, the corresponding set of poles is not split in subsets by the contours. Therefore, the result of Eq.(\ref{2_MB_F1}) can be directly checked using the original version of our \texttt{MBConicHulls.wl} \textit{Mathematica} package (\textit{i.e.} without explicitly fixing the contours), as the package is based on this assumption. Doing this exercise, one obtains from the package that there are 8 different conic hulls associated to this MB integral and that these lead to 5 different series representations. The simplest of the latter is the first one, obtained using the \texttt{MBResolve[,1]} and \texttt{EvaluateSeries[,1]} commands of the package. It gives the well-known double series representation of $F_1$ and its overall factor written in Eq.(\ref{2_MB_F1}) above.

We now want to compute the MB integral of Eq.(\ref{2_MB}) in the less trivial situation where $c_1=\frac{7}{3}$ and $c_2=-\frac{3}{2}$.

In this case, the first and fifth gamma functions of the numerator of the MB integrand in Eq.(\ref{2_MB}) have arguments with negative real parts. Therefore, their sets of poles are now split by the contours, as can be seen in Fig. \ref{Singular} \textit{Left} (the point $(c_1,c_2)=(\frac{7}{3},-\frac{3}{2})$ is located between singular lines having the same color: two are yellow and two are green). Therefore, in order to apply our package for the evaluation of the corresponding MB integral, we have to find a way to transform these gamma functions in order that the real part of their arguments become positive for $c_1=\frac{7}{3}$ and $c_2=-\frac{3}{2}$. This can be done at the cost of introducing more gamma functions in the integrand, by using the generalized reflection formula:
\begin{align}
\Gamma(z-n)=\frac{\Gamma(z)\Gamma(1-z)(-1)^n}{\Gamma(n+1-z)}\label{reflection}
\end{align}
Indeed, rewriting the first gamma function as
\begin{align}\label{firstReflection}
\Gamma(-z_1)=\frac{\Gamma(3-z_1)\Gamma(-2+z_1)(-1)^3}{\Gamma(1+z_1)}
\end{align}
and the fifth one as 
\begin{align}\label{secondReflection}
\Gamma\left(\frac{3}{5}+z_2\right)=-\frac{\Gamma(-\frac{3}{5}-z_2)\Gamma(\frac{8}{5}+z_2)}{\Gamma(\frac{2}{5}-z_2)}
\end{align}
one sees that for $c_1=\frac{7}{3}$ and $c_2=-\frac{3}{2}$ both gamma functions in the numerator of the RHS of Eq.(\ref{firstReflection}) and Eq.(\ref{secondReflection}) will now have arguments with positive real parts (in contrary with the LHS).

This allows us to write Eq.(\ref{2_MB}) in the following equivalent way
\begin{align} \label{2_MB2}
    I (x_1,x_2) = \int\limits_{c_1-i \infty}^{c_1+i \infty} \frac{ \text{d} z_1}{2 \pi i} \int\limits_{c_2-i \infty}^{c_2+i \infty}\frac{ \text{d} z_2}{2 \pi i}\,\, (-x_1)^{z_1} &(-x_2)^{z_2} \frac{\Gamma(3-z_1)\Gamma(-2+z_1)}{\Gamma(1+z_1)}\Gamma(-z_2) \nonumber\\
    &\times   \frac{ \Gamma(\frac{3}{7}+z_1+z_2)\Gamma(\frac{2}{3}+z_1)}{\Gamma(\frac{1}{2}+z_1+z_2)}\frac{\Gamma(-\frac{3}{5}-z_2)\Gamma(\frac{8}{5}+z_2)}{\Gamma(\frac{2}{5}-z_2)} 
\end{align}
where now all the gamma functions in the numerator have arguments with positive real parts for the chosen contours $c_1=\frac{7}{3}$ and $c_2=-\frac{3}{2}$. 

\begin{figure}[h]
\centering
\includegraphics[width=8cm, height=8cm]{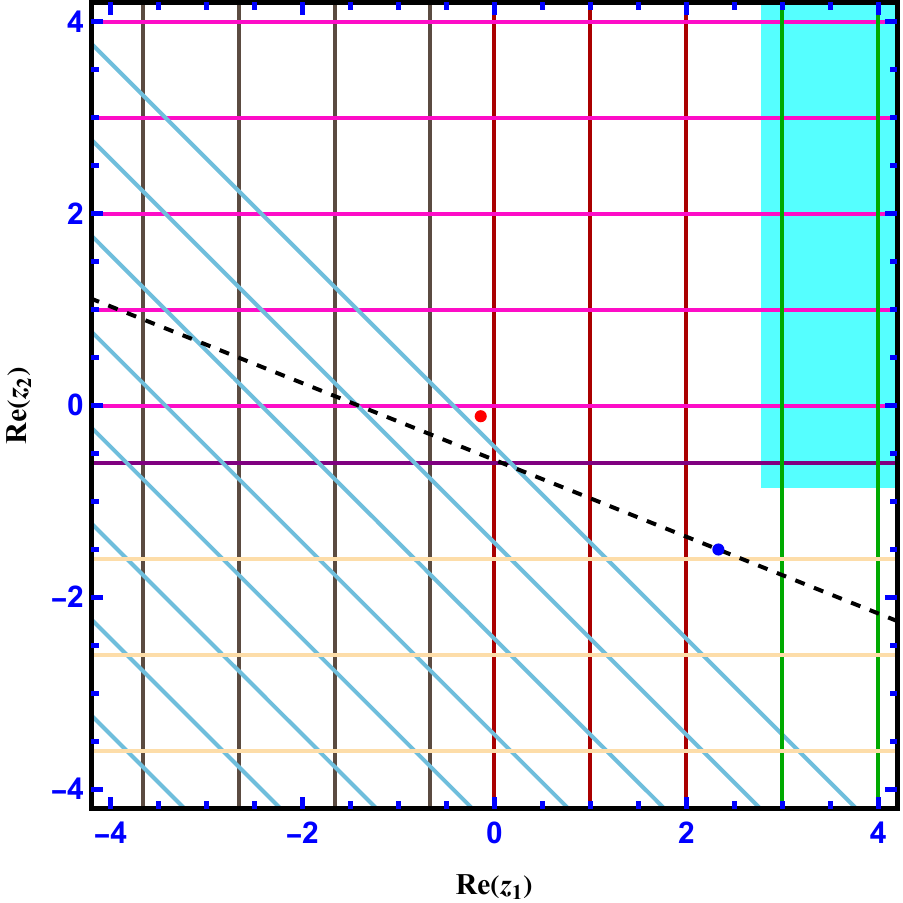}
\caption{Cone (shaded area in light blue) associated with the series representation of Eq.(\ref{2_MB}) computed in the text. We refer the reader to \cite{Friot:2011ic} for details about the derivation of the cone using the dashed black line.\label{Cone}}
\end{figure}

As, for $c_1=\frac{7}{3}$ and $c_2=-\frac{3}{2}$, this MB integral is now satisfying the main constraint for the use of the original version of our \texttt{MBConicHulls.wl} package  (see Fig. \ref{Singular} \textit{Right} where the point $(c_1,c_2)=(\frac{7}{3},-\frac{3}{2})$ is not anymore located between singular lines having the same color), it can be computed with the latter (\textit{i.e} without specifying the contours of integrations). However, there are now 15 different conic hulls, which lead to 5 possible series representations for the MB integral of Eq.(\ref{2_MB2}) and the package gives, for the first series representation, the following result:
\begin{align}
I(x_1,x_2)\underset{\vert x_1\vert<1\wedge\vert x_2\vert<1}=&(-x_1)^3\sum_{n_1=0,n_2=0}^{\infty}\frac{\Gamma(\frac{11}{3}+n_1)\Gamma(-\frac{3}{5}-n_2)\Gamma(\frac{8}{5}+n_2)\Gamma(\frac{24}{7}+n_1+n_2)}{\Gamma(4+n_1)\Gamma(\frac{2}{5}-n_2)\Gamma(\frac{7}{2}+n_1+n_2)}x_1^{n_1}\frac{x_2^{n_2}}{n_2!}\nonumber\\
&+(-x_1)^3(-x_2)^{-\frac{3}{5}}\Gamma\left(\frac{3}{5}\right)\sum_{n_1=0}^{\infty}\frac{\Gamma(\frac{11}{3}+n_1)\Gamma(\frac{99}{35}+n_1)}{\Gamma(4+n_1)\Gamma(\frac{29}{10}+n_1)}x_1^{n_1}\label{result}
\end{align}
One can check that this result is correct by directly computing Eq.(\ref{2_MB}) with $c_1=\frac{7}{3}$ and $c_2=-\frac{3}{2}$ using the computational approach of \cite{Friot:2011ic} (see also \cite{Passare:1996db,TZ}). Indeed, it can be seen from Fig. \ref{Cone} that the cone corresponding to this series representation has two different sets of singular points from which Eq.(\ref{result}) can be reobtained.
The first set is $(z_1,z_2)=(3+n_1,n_2)$ which gives the contribution
\begin{align}
I_1=-(-x_1)^3\frac{\Gamma(\frac{24}{7})\Gamma( \frac{11}{3})\Gamma( \frac{3}{5})}{\Gamma( \frac{7}{2})\Gamma(4)}{F}{}^{1:2;1}_{1:1;0}
  \left[
   \setlength{\arraycolsep}{0pt}
   \begin{array}{c@{{}:{}}c@{;{}}c}
  \frac{24}{7} & \frac{11}{3},1 & \frac{3}{5}\\[1ex]
   \frac{7}{2} & 4 & -
   \end{array}
   \;\middle|\;
x_1, x_2
 \right]
\end{align}
where ${F}{}^{1:2;1}_{1:1;0}$ is a Kamp\'e de F\'eriet double hypergeometric series \cite{KdF,Srivastava}. $I_1$ matches with the first term of Eq.(\ref{result}) once Eq.(\ref{reflection}) is used in the latter. The second set is $(z_1,z_2)=(3+n_1,-\frac{3}{5})$, whose associated residues give
\begin{align}
I_2=(-x_1)^3(-x_2)^{-\frac{3}{5}}\Gamma\left(\frac{3}{5}\right)\frac{\Gamma(\frac{99}{35})\Gamma(\frac{11}{3})}{\Gamma(4)\Gamma(\frac{29}{10})}{}_3F_2\left(\frac{99}{35},\frac{11}{3},1;4,\frac{29}{10};x_1\right)
\end{align}
which is equal to the second term of Eq.(\ref{result}).

We have seen above, as a motivating example, how to treat a simple case of multifold MB integral with arbitrary straight contours, by hand. We now show how the same calculations can be performed in an automatic way using the new version of our \texttt{MBConicHulls.wl} package that we have implemented and which now allows the user to choose, if needed, arbitrary straight contours of integration for multifold MB integrals (details about some of the commands that are used in this calculations can be found in \cite{Ananthanarayan:2020fhl}).

For this, we first load in a \textit{Mathematica} notebook the package as follows:
\medskip

\texttt{In[1]:SetDirectory[NotebookDirectory[]];}

\texttt{In[2]:=<<MBConicHulls.wl;}
\medskip

\noindent where we assume that the \texttt{MBConicHulls.wl} package is kept in the same directory as the notebook. The \texttt{MultivariateResidues.m} package \cite{Larsen:2017aqb} also has to be put in this directory since it will be called internally by \texttt{MBConicHulls.wl}.

\noindent Then we input our MB integral with the above chosen contours, using the \texttt{MBRep[]} command, as follows:
\medskip

\texttt{In[3]:MBRepOut=MBRep[1,}$\mathtt{\{z_1}\mathtt{\rightarrow}\frac{\texttt{7}}{\texttt{3}}\mathtt{,z_2}\mathtt{\rightarrow-\frac{\texttt{3}}{\texttt{2}}\}, \{-z_1,-z_2,\frac{3}{7}+z_1+z_2,\frac{2}{3}+z_1,\frac{3}{5}+z_2\}}\\ \mathtt{\{\frac{1}{2}+z_1+z_2\}}$\texttt{];}
\medskip

\noindent which gives as an output \medskip

\texttt{Straight\ contours :} $\{\mathtt{Re(z_\texttt{1})\rightarrow\frac{\texttt{7}}{\texttt{3}},Re(z_\texttt{2})\rightarrow-\frac{\texttt{3}}{\texttt{2}}}\}$
\medskip

\noindent One can then find the series representations of \texttt{MBRepOut} using the \texttt{ResolveMB[]} command. Let us compute the first of the 5 possible series representations by the instruction
\medskip

\texttt{In[4]:ResolveMBOut=ResolveMB[MBrepOut,1];}
\medskip

\noindent where \texttt{1} indicates that we are interested in only the first series solution. This gives
\medskip

\texttt{Degenerate case with 15 conic hulls}

\texttt{Series\ solution\ 1:: Intersecting conic hulls $\mathtt{\{C_{1,2},C_{2,5}\}}$. The set of poles is :: $\{\mathtt{(3+ n_2, n_1),(3+ n_1,-\frac{3}{5}- n_2)}\}$ with master series characteristic list and variables $\{\mathtt{(n_1, n_2),(-x_1,-x_2)}\}$.}
\medskip

\noindent From the informations given by the package about the master series, it is clear that the convergence region of the first series representation is the same as the one indicated under the equal sign of Eq.(\ref{result}).

\noindent Therefore, in order to obtain the explicit expression of this first series representation, one inputs
\medskip

\texttt{In[5]:EvaluateSeries[ResolveMBOut,1];}
\medskip

\noindent which indeed exactly gives Eq.(\ref{result}) (we do not reproduce the final result here). 

Let us now consider a non-trivial physical application of our approach.

\section{The one-loop massless pentagon in $D=4-2\epsilon$ \label{section2}}

In this section, we apply our method to the computation of the dimensionally regularized massless one-loop pentagon integral in general kinematics and $D=4-2\epsilon$ space-time dimensions. Indeed, in the MB representation approach, the $\epsilon$-expansion of this object obtained from the \textit{Mathematica} packages \texttt{MBresolve.m} \cite{Smirnov:2009up} or \texttt{MB.m} \cite{Czakon:2005rk} requires the computation of a combination of eleven MB integrals (up to 4-fold) with straight contours, which gives us the opportunity to test the new features of our \texttt{MBConicHulls.wl} package in a non-trivial physical case. The results will be checked by direct numerical integration of the involved multifold MB integrals, as well as by computing the pentagon using \texttt{FIESTA} \cite{Smirnov:2015mct}. We will also compare our results with those obtained from the package \texttt{MBsums.m} \cite{Ochman:2015fho}.

The MB representation of the pentagon was first given in \cite{Bern:2006vw} for general powers of the propagators and $D$ dimensions, and it was studied in detail  in \cite{DelDuca:2009ac}, in the limit of multi-Regge kinematics, for unit propagator powers and $D=6-2\epsilon$. In the present work, we also focus on the resonant unit powers of propagators case, but for $D=4-2\epsilon$ and general kinematics (see \cite{Kniehl:2010aj} for investigations in $D$ dimensions with another approach). In this situation, the MB representation of the pentagon reads (using the same notation as in \cite{DelDuca:2009ac}):
\begin{align}
I^{4-2\epsilon}_5(1,1,1,1,1;Q_i^2)&=\frac{-e^{\gamma_E\epsilon} (-s)^{-3-\epsilon}}{\Gamma(-1-2\epsilon)}\,I_{\mathrm{4MB}}
\label{MBrepPentagon1}
\end{align}
where
\begin{align}
I_{\mathrm{4MB}}&= \int\limits_{-i \infty}^{+i \infty} \frac{ \text{d} z_1}{2 \pi i} \cdots \int\limits_{-i \infty}^{+i \infty}\frac{ \text{d} z_4}{2 \pi i}\,\, u_1^{z_1}u_2^{z_2} u_3^{z_3} u_4^{z_4} \prod\limits_{i=1}^{4} \Gamma(-z_i)\Gamma(1+z_{12})\Gamma(-2-\epsilon-z_{123}) \nonumber\\
&\times\Gamma(1+z_{23})\Gamma(-2-\epsilon-z_{234})\Gamma(1+z_{34})\Gamma(3+\epsilon+z_{1234})
\label{MBrepPentagon}
\end{align}
where $u_1=\frac{s_2}{s}, u_2=\frac{t_1}{s}, u_3=\frac{t_2}{s}$ and $u_4=\frac{s_1}{s}$. We use the shorthand notation $z_{ij\,\cdots}=z_i+z_j+\cdots$ and in the following we work in the Euclidean region where $s, s_1, s_2, t_1$ and $t_2$ are negative.

It is implicit, in the notation of this 4-fold MB integral, that the contours of integrations are non-straight, and this integral represents the pentagon integral when the contours satisfy the usual assumption that the set of poles of each of the gamma functions of the integrand (there is no denominator in this example) is not split into different subsets. We note here that it is possible, for this integral, to find straight contours having this property, but not for $\epsilon=0$ (for instance if we choose the contours such that $\Re{(z_i)}=-0.1$ for $i=1,...,4$ then we must have $-3.4<\epsilon<-1.7$).

At this point we remark that as the MB integral in Eq.(\ref{MBrepPentagon}) has non-straight contours which satisfy the condition for using the original version of \texttt{MBConicHulls.wl}, it is a simple exercise (although this is a non-trivial resonant case \cite{Ananthanarayan:2020fhl}) to derive with the package some of its various series representations for general kinematics and general $\epsilon$. We performed this exercise in the \texttt{Pentagon.nb} \textit{Mathematica} notebook which can be downloaded from \cite{git}. One of the 70 possible series representations that can be extracted from a direct evaluation of the pentagon MB integral representation is also given explicitly in Appendix B, see Eq.(\ref{S}). To our knowledge, this expression is new. We have checked it numerically with \texttt{FIESTA} (as well as two other series representations that we do not give here) for 3 Euclidean points and we got perfect agreement when choosing a small value of $\epsilon$ and some values of the $u_i$ such that Eq.(\ref{S}) converges (for instance $u_1=0.0001, u_2=0.001, u_3=0.01$ and  $u_4=0.1$). It is also possible, fixing for instance $\epsilon=-2$, to use \texttt{MBsums.m} in order to compute Eq.(\ref{MBrepPentagon}) if one chooses the straight contours mentioned above ($\Re{(z_i)}=-0.1$ for $i=1,...,4$) and, therefore, one can compare the obtained results with ours. We found a complete numerical agreement between both analytic expressions, although the number of terms given by our approach is 14 while \texttt{MBsums.m} gives a less compact expression involving 48 terms (we recall however that the expression given by \texttt{MBsums.m} can only be used for a particular numerical check, because it is not valid for $\epsilon\neq-2$). We also got agreement with the direct numerical integration of Eq.(\ref{MBrepPentagon}) obtained (still for these particular values of the contours and $\epsilon$) with the help of the \texttt{MBintegrate[]} command of \texttt{MB.m}. 

While the results discussed above for general $\epsilon$ are interesting (in particular because, to our knowledge, our \texttt{MBConicHulls.wl} package is the only available tool for such analytic calculations with non-straight contours), in the rest of this section we will focus on first performing the $\epsilon$-expansion of the MB integral using \texttt{MBresolve.m} (or \texttt{MB.m}). Indeed, this leads to the evaluation of several MB integrals with straight contours on which we can apply the techniques discussed earlier in this paper and demonstrate the efficiency of our approach in deriving analytic solutions.

As shown in the \texttt{Pentagon.nb} notebook, the result of the $\epsilon$-expansion given by the package \texttt{MBresolve.m} is a sum of a constant term, five 1-fold, three 2-fold, two 3-fold and one 4-fold MB integrals with straight contours, all of them \textit{not} satisfying the constraint of positivity of the real part of their arguments for the contours given by this package. Therefore, this provides an ideal non-trivial situation to test our computational technique and its implementation in \texttt{MBConicHulls.wl}. We have listed all these (resonant) MB integrals and the constant term in Appendix A and a detailed analysis can be found in \texttt{Pentagon.nb} where we derived series representations for each of the MB integrals and cross-checked them with direct numerical integration using \texttt{MB.m}. Alternatively, we numerically cross-checked our results using the corresponding analytic series solutions derived from \texttt{MBsums.m}, although the latter often yields more lengthy series solutions than our approach. We also got a perfect numerical matching with \texttt{FIESTA}.

At the sample points where we performed the numerical checks in this section, the series representations were rapidly converging and the corresponding numerical results were obtained in a few seconds by summing 10 to 15 terms for each of the summation indices.

As a final (side) remark, we note that the first version of \texttt{MBConicHulls.wl} could not evaluate $1$-fold MB integrals, therefore we upgraded our package so that the new version can now also evaluate this class of integrals (with straight or non-straight contours). Hence, the checks mentioned above have been performed by including, in the automatic evaluation, the $1$-fold MB integrals of the pentagon given in Eqs.(\ref{intMB1fold1})-(\ref{intMB1fold5}).

\section{Conclusions}

In this work, we have shown how the computational method of \cite{Ananthanarayan:2020fhl}, dedicated to the evaluation of multifold MB integrals, can be adapted to deal with the case of arbitrary straight contours of integration. We have detailed, on a 2-fold example, how this can be achieved by using simple transformations of the MB integrand and we have automated this procedure in a new version of the \texttt{MBConicHulls.wl} package (see \cite{git}).
This package has been tested on several examples, including the non-trivial computation of the $\epsilon$-expansion of the (resonant) one-loop massless pentagon in $D=4-2\epsilon$ with unit propagator powers, which involves a combination of eleven MB integrals with straight contours (and up to 4-fold). An excellent agreement has been obtained when comparing the evaluation of the pentagon from the analytic computation of these integrals, derived with \texttt{MBConicHulls.wl}, with the direct numerical evaluation of this Feynman integral with the help of \texttt{FIESTA} \cite{Smirnov:2015mct} (as well as with the numerical evaluation of each of the eleven MB integrals separately using \texttt{MB.m} \cite{Czakon:2005rk}). We have also compared our results with those derived from \texttt{MBsums.m} \cite{Ochman:2015fho}, an alternative \textit{Mathematica} package available for the analytic computation of MB integrals with straight contours. Although we got complete numerical agreement with the expressions derived from the latter, our approach gives more compact results, confirming the concluding remarks of \cite{Ananthanarayan:2020fhl}. This is probably due to the fact that \texttt{MBsums.m} computes the multifold MB with straight contours iteratively, one integral after the other, whereas our package is based on a non-iterative technique based on multidimensional residue theory. A deeper analysis is needed to better understand the difference between the results obtained from these two methods, which is left for future investigations. 

We conclude this work by mentioning that, although our \texttt{MBConicHulls.wl} package is, to our knowledge, the unique tool that can treat the automatic analytic calculations of multifold MB integrals with both straight and non-straight contours\footnote{Indeed,  the \texttt{MBsums.m} package is restricted to the calculation of MB integrals whose contours are straight lines parallel to the imaginary axis in the complex planes of the integration variables and, therefore, cannot be used to compute multifold MB integrals with non-fixed values of the Pochhammer parameters (for which one cannot have straight contours) or when these parameters are fixed, but in such a way that the contours are non-straight. Both of these situations can be treated by \texttt{MBConicHulls.wl}.}, which is of considerable interest in the theory of MB integrals, as well as for the study of multivariable hypergeometric functions, their transformation theory, and for other domains where these integrals appear, there is still one progress that has to be achieved in order that our computational technique and package can be more widely used for phenomenological applications in particle physics. Indeed, in its present version  \texttt{MBConicHulls.wl} can be used only when the number of $x_i$ in Eq.(\ref{N_MB}) is equal than the number of integration variables $z_i$. It would thus be important to find a way to apply the method in the case where this number of $x_i$ is smaller than the number of folds of the MB integrals (and when Barnes lemmas cannot be applied to lower the number of folds), because this is a common situation met when computing Feynman diagrams in particle physics. This interesting problem will have to be addressed in the future.

\section{Appendix A }
After $\epsilon$-expansion up to order $\epsilon^0$, the 4-fold MB integral in Eq. \eqref{MBrepPentagon} can be written as
\begin{align}
I^{\mathrm{4MB}}=J^{0}+\sum_{i=1}^{5} J^{\mathrm{1MB}}_{i}+\sum_{i=1}^{3}J^{\mathrm{2MB}}_{i}+\sum_{i=1}^{2}J^{\mathrm{3MB}}_{i}+J^{\mathrm{4MB}}
\label{Pentagon_Exp}
\end{align}

where Strategy A has been used through \texttt{MBresolve.m}  \cite{Smirnov:2009up} (\texttt{MB.m} \cite{Ochman:2015fho} gives a similar result).

All the terms on the RHS, except $J^0$, involve MB integrals with straight contours which split the sets of poles of some of the numerator gamma functions in different subsets and thus for which the method presented in this paper is necessary in order to apply the conic hull approach to compute these integrals analytically. We emphasize that these integrals are non-trivial as they belong to the resonant class \cite{Ananthanarayan:2020fhl}. A detailed analysis can be found in the \texttt{Pentagon.nb} file. However, for completeness, we present each of the terms in the RHS of Eq. \eqref{Pentagon_Exp} below along with the number of conic hulls associated and number of series solutions obtained from the extended conic hull approach.

\begin{itemize}

\item The integral-less first term on the RHS of Eq. \eqref{Pentagon_Exp}  is
\begin{align}
J_0&={u_1}^{-\epsilon -1} {u_2}^{-\epsilon -1} {u_3}^{\epsilon } {u_4}^{-\epsilon -1} \Gamma (-2 \epsilon
   -1) \Gamma (-\epsilon )^2 \Gamma (\epsilon +1)^3
   
   \nonumber\\&
   
   -{u_1}^{-\epsilon } {u_2}^{-\epsilon -1}
   {u_3}^{\epsilon } {u_4}^{-\epsilon -1} \Gamma (1-\epsilon ) \Gamma (-2 \epsilon ) \Gamma
   (-\epsilon ) \Gamma (\epsilon ) \Gamma (\epsilon +1)^2
   
   \nonumber\\&
      
   +{u_1}^{-\epsilon -1} {u_2}^{\epsilon
   } {u_3}^{-\epsilon -1} {u_4}^{-\epsilon -1} \Gamma (-2 \epsilon -1) \Gamma (-\epsilon )^2
   \Gamma (\epsilon +1)^3

\nonumber\\&   
   
   +{u_1}^{-\epsilon -1} {u_2}^{\epsilon } {u_3}^{-\epsilon -1} {u_4}^{\epsilon }
   \Gamma (-2 \epsilon -1) \Gamma (-\epsilon )^2 \Gamma (\epsilon +1)^3
   
\nonumber\\&
   
   +{u_1}^{\epsilon }
   {u_2}^{-\epsilon -1} {u_3}^{-\epsilon -1} {u_4}^{\epsilon } \Gamma (-2 \epsilon -1) \Gamma
   (-\epsilon )^2 \Gamma (\epsilon +1)^3
   
   \nonumber\\&
   
   +{u_2}^{-\epsilon -1} {u_3}^{\epsilon } {u_4}^{-\epsilon
   -1} \Gamma (-\epsilon -1) \Gamma (-\epsilon )^2 \Gamma (\epsilon +1)^2
\end{align}

\item 1-fold MB integrals: 

There are five 1-fold integrals. These integrals contribute from the $\epsilon^{-1}$ term of the $\epsilon$-expansion, therefore, we keep their $\epsilon$ dependency explicit as we perform the $\epsilon$-expansion on the series solutions derived using our approach.

\begin{equation}\label{intMB1fold1}
J^{\mathrm{1MB}}_{1}= {u_2}^{-\epsilon -1} {u_3}^{\epsilon } {u_4}^{-\epsilon -1} \Gamma (-\epsilon ) \Gamma^2
   (\epsilon +1) \int\limits_{\frac{4}{5}-i \infty}^{\frac{4}{5}+i \infty} \frac{ \text{d} z_1}{2 \pi i} \, {u_1}^{z_1} \, \Gamma(-z_1) \Gamma(1+z_1)\Gamma(-\epsilon+z_1)\Gamma(-1-\epsilon-z_1)
\end{equation}

\begin{equation}
J^{\mathrm{1MB}}_{2}= {u_3}^{-\epsilon -1} {u_4}^{\epsilon } {u_1}^{-1} \Gamma (-\epsilon ) \Gamma^2
   (\epsilon +1) \int\limits_{-\frac{4}{5}-i \infty}^{-\frac{4}{5}+i \infty} \frac{ \text{d} z_2}{2 \pi i} \left(\frac{{u_2}}{{u_1}}\right)^{z_2} \Gamma(-z_2) \Gamma(1+z_2)\Gamma(-\epsilon+z_2)\Gamma(-1-\epsilon-z_2)
\end{equation}

\begin{equation}
J^{\mathrm{1MB}}_{3}= {u_1}^{-\epsilon -1} {u_4}^{-1-\epsilon} {u_2}^{-1} \Gamma (-\epsilon ) \Gamma^2
   (\epsilon +1) \int\limits_{-\frac{4}{5}-i \infty}^{-\frac{4}{5}+i \infty} \frac{ \text{d} z_3}{2 \pi i} \left(\frac{{u_3}}{{u_2}}\right)^{z_3} \Gamma(-z_3) \Gamma(1+z_3)\Gamma(-\epsilon+z_3)\Gamma(-1-\epsilon-z_3)
\end{equation}

\begin{equation}
J^{\mathrm{1MB}}_{4}= {u_2}^{-\epsilon -1} {u_3}^{-1} \Gamma (-\epsilon ) \Gamma
   (\epsilon +1) \int\limits_{-\frac{4}{5}-i \infty}^{-\frac{4}{5}+i \infty} \frac{ \text{d} z_4}{2 \pi i} \left(\frac{{u_4}}{{u_3}}\right)^{z_4} \Gamma(-z_4) \Gamma(1+z_4)\Gamma(+z_4)\Gamma(-1-\epsilon-z_4)
\end{equation}

\begin{equation}\label{intMB1fold5}
J^{\mathrm{1MB}}_{5}= {u_1}^{-\epsilon -1} {u_2}^{\epsilon} {u_3}^{-1-\epsilon} \Gamma (-\epsilon ) \Gamma^2
   (\epsilon +1) \int\limits_{-\frac{4}{5}-i \infty}^{-\frac{4}{5}+i \infty} \frac{ \text{d} z_4}{2 \pi i} \, {u_4}^{z_4} \, \Gamma(-z_4) \Gamma(1+z_4)\Gamma(-\epsilon+z_4)\Gamma(-1-\epsilon-z_4)
\end{equation}

$J^{\mathrm{1MB}}_{1}$ has $5$ conic hulls while the other four integrals have $6$ conic hulls each. All these integrals have $2$ possible series solutions that are analytic continuations of each other.
\bigskip

\item 2-fold MB integrals: 

There are three 2-fold integrals. These integrals and all the remaining integrals contribute from the $\epsilon^0$ term of the $\epsilon$-expansion. Therefore, we set $\epsilon=0$ at the integrand level before evaluating their series solutions.

\begin{align}
J^{\mathrm{2MB}}_{1}= \frac{1}{{u_2} {u_3}} \int\limits_{\frac{1}{10}-i \infty}^{\frac{1}{10}+i \infty} \frac{\text{d} z_1}{2 \pi i} \int\limits_{-\frac{1}{15}-i \infty}^{-\frac{1}{15}+i \infty} \frac{\text{d} z_4}{2 \pi i} \, & {u_1}^{z_1} \left(\frac{{u_4}}{{u_3}} \right)^{z_4} \, \Gamma \left(-z_1\right)  \Gamma \left(-z_4\right) \Gamma \left(z_1\right)  \Gamma \left(z_1+1\right) \nonumber \\ & \times 
   \Gamma
   \left(-z_4-1\right)  \Gamma \left(z_4+1\right) \Gamma\left(z_4-z_1\right)
\end{align}
has $26$ associated conic hulls and $6$ series solutions.

\begin{align}
J^{\mathrm{2MB}}_{2}= \frac{1}{{u_1} {u_3}} \int\limits_{-\frac{7}{5}-i \infty}^{-\frac{7}{5}+i \infty} \frac{\text{d} z_2}{2 \pi i} \int\limits_{-\frac{1}{2}-i \infty}^{-\frac{1}{2}+i \infty} \frac{\text{d} z_4}{2 \pi i} \, & \left(\frac{{u_2}}{{u_1}} \right)^{z_2} \left(\frac{{u_4} {u_1}}{{u_3}} \right)^{z_4} \Gamma \left(-z_2\right) \Gamma \left(-z_4\right)  \Gamma \left(-z_2-1\right)  \nonumber \\ & \times
  \Gamma \left(z_2-z_4\right)\Gamma\left(z_2-z_4+1\right)  \Gamma \left(z_4\right) \Gamma^2
   \left(z_4+1\right)
\end{align}

has $31$ associated conic hulls and $5$ series solutions.

\begin{align}
J^{\mathrm{2MB}}_{3}= \frac{1}{{u_1} {u_2}} \int\limits_{-\frac{1}{3}-i \infty}^{-\frac{1}{3}+i \infty} \frac{\text{d} z_3}{2 \pi i} \int\limits_{-\frac{2}{5}-i \infty}^{-\frac{2}{5}+i \infty} \frac{\text{d} z_4}{2 \pi i} \, & \left(\frac{{u_3}}{{u_2}} \right)^{z_3} {u_4}^{z_4} \, \Gamma \left(-z_3\right) \Gamma \left(-z_4\right) \Gamma \left(-z_3-1\right)   \nonumber \\ & \times
     \Gamma \left(z_3+1\right) \Gamma
   \left(-z_4-1\right)\Gamma \left(z_4+1\right) \Gamma
   \left(z_3+z_4+1\right)
\end{align}

has $24$ associated conic hulls and $5$ series solutions.
\bigskip

\item 3-fold MB integrals: 

There are two 3-fold integrals:

\begin{align}
J^{\mathrm{3MB}}_{1}= &\frac{1}{{u_2}^2}  \int\limits_{-\frac{2}{5}-i \infty}^{-\frac{2}{5}+i \infty} \frac{\text{d} z_1}{2 \pi i} \int\limits_{-\frac{4}{5}-i \infty}^{-\frac{4}{5}+i \infty} \frac{\text{d} z_3}{2 \pi i}  \int\limits_{-\frac{11}{10}-i \infty}^{-\frac{11}{10}+i \infty} \frac{\text{d} z_4}{2 \pi i} \,  {u_1}^{z_1}\left(\frac{{u_3}}{{u_2}} \right)^{z_3}\left(\frac{{u_4}}{{u_2}} \right)^{z_4}
 \, \Gamma \left(-z_1\right) \Gamma \left(-z_3\right) \Gamma \left(-z_4\right)
 
 \nonumber \\ & \times  \Gamma \left(z_1+1\right)  \Gamma \left(-z_4-1\right) \Gamma \left(z_1-z_3-z_4-1\right) 
   \Gamma \left(z_4-z_1\right) \Gamma \left(z_3+z_4+1\right) \Gamma
   \left(z_3+z_4+2\right)
\end{align}

\begin{align}
J^{\mathrm{3MB}}_{2}= &\frac{1}{{u_1}^2}  \int\limits_{-\frac{3}{5}-i \infty}^{-\frac{3}{5}+i \infty} \frac{\text{d} z_2}{2 \pi i} \int\limits_{-\frac{11}{10}-i \infty}^{-\frac{11}{10}+i \infty} \frac{\text{d} z_3}{2 \pi i} \int\limits_{-\frac{1}{2}-i \infty}^{-\frac{1}{2}+i \infty} \frac{\text{d} z_4}{2 \pi i} \, \left(\frac{{u_2}}{{u_1}}\right)^{z_2}\left(\frac{{u_3}}{{u_1}} \right)^{z_3}{u_4}^{z_4} \, \Gamma \left(-z_2\right) \Gamma \left(-z_3\right) \Gamma \left(-z_4\right) \nonumber \\ & \times  \Gamma \left(-z_3-1\right) \Gamma
   \left(z_2+z_3+1\right) \Gamma \left(z_2+z_3+2\right) \Gamma
   \left(-z_2-z_3-z_4-2\right)  \Gamma \left(z_4+1\right)
   \Gamma \left(z_3+z_4+1\right)
\end{align}

Both the above 3-fold MB integrals have $91$ associated conic hulls and $20$ associated series solutions.
\bigskip

\item 4-fold MB integrals: 

There is only one 4-fold integral

\begin{align}
J^{\mathrm{4MB}}&=\int\limits_{-\frac{1}{5}-i \infty}^{-\frac{1}{5}+i \infty} \frac{ \text{d} z_1}{2 \pi i} \int\limits_{-\frac{1}{2}-i \infty}^{-\frac{1}{2}+i \infty}\frac{ \text{d} z_2}{2 \pi i}\int\limits_{-\frac{2}{5}-i \infty}^{-\frac{2}{5}+i \infty}\frac{ \text{d} z_3}{2 \pi i} \int\limits_{-\frac{2}{5}-i \infty}^{-\frac{2}{5}+i \infty}\frac{ \text{d} z_4}{2 \pi i}\,\, {u_1}^{z_1}{u_2}^{z_2}{u_3}^{z_3}{u_4}^{z_4} \prod\limits_{i=1}^{4} \Gamma(-z_i)\,\,\Gamma(1+z_{12}) 
\nonumber\\&\times
\Gamma(-2-z_{123}) \Gamma(1+z_{23})\Gamma(-2-z_{234})\Gamma(1+z_{34})\Gamma(3+z_{1234})
\end{align}
with $245$  associated conic hulls and $70$ series solutions.
\end{itemize}

\section{Appendix B}
In this Appendix, we present one series representation (denoted as $S$ below) of the pentagon diagram obtained by applying the conic hull method on the resonant $4$-fold MB integral in Eq.\eqref{MBrepPentagon} for general values of $\epsilon$. 69 other series representations can be easily derived from this MB integral using the \texttt{MBConichulls.wl} package.

This series representation $S$ is a sum of $14$ terms
\begin{equation}\label{S}
S=\sum_{i=1}^{14} S_i 
\end{equation}
where 
\begin{align}
S_1=& \sum_{n_i=0}^{\infty} \Gamma \left(n_1+n_2+1\right) \Gamma \left(n_2+n_3+1\right) \Gamma
   \left(n_3+n_4+1\right) \Gamma \left(-\epsilon -n_1-n_2-n_3-2\right)\nonumber \\ & \times
    \Gamma \left(-\epsilon -n_2-n_3-n_4-2\right) \Gamma \left(\epsilon
   +n_1+n_2+n_3+n_4+3\right) \frac{(-u_1)^{n_1} (-u_2)^{n_2} (-u_3)^{n_3} (-u_4)^{n_4}}{n_1! \, n_2! \, n_3! \, n_4!} 
\end{align}

\begin{align}
S_2=& u_4^{-2-\epsilon} \sum_{n_i=0}^{\infty} \Gamma \left(n_1+n_2+1\right) \Gamma \left(n_2+n_3+1\right) \Gamma \left(n_1+n_4+1\right) \Gamma \left(-\epsilon -n_1-n_2-n_3-2\right)
\nonumber \\ & \times
 \Gamma \left(\epsilon +n_2+n_3-n_4+2\right) \Gamma \left(-\epsilon -n_2+n_4-1\right)
 \frac{(-u_1)^{n_1} (-u_2/u_4)^{n_2} (-u_3/u_4)^{n_3} (-u_4)^{n_4}}{n_1! \, n_2! \, n_3! \, n_4!} 
\end{align}

\begin{align}
S_3=& u_3^{-2-\epsilon} \sum\limits_{n_1 > n_3 +n_4} \Gamma \left(n_1+n_2+1\right) \Gamma \left(n_1-n_3-n_4\right) \Gamma \left(n_3+n_4+1\right) \Gamma \left(\epsilon +n_1+n_2-n_4+2\right)
\nonumber \\ & \times
 \Gamma
   \left(-\epsilon -n_1+n_4-1\right) \Gamma \left(-\epsilon -n_1-n_2+n_3+n_4-1\right)
 \frac{(-u_1/u_3)^{n_1} (-u_2/u_3)^{n_2} (-u_4)^{n_3} (-u_3)^{n_4}}{n_1! \, n_2! \, n_3! \, n_4!} 
\end{align}

\begin{align}
S_4=& -u_3^{-2-\epsilon} \sum\limits_{n_1 \leq n_3 +n_4} 
\frac{\Gamma \left(n_1+n_2+1\right) \Gamma \left(n_3+n_4+1\right) \Gamma \left(\epsilon +n_1+n_2-n_4+2\right) \Gamma \left(-\epsilon -n_1+n_4-1\right)}{\Gamma(1-n_1+n_3+n_4)}
\nonumber \\ & \times
   \Gamma \left(-\epsilon -n_1-n_2+n_3+n_4-1\right) \bigg(\psi\left(-\epsilon -n_1-n_2+n_3+n_4-1\right)-\psi\left(n_3+1\right)
\nonumber \\ & 
   +\psi\left(n_3+n_4+1\right)-\psi\left(-n_1+n_3+n_4+1\right)+\log \left(u_4\right)\bigg)
 \frac{(u_1/u_3)^{n_1} (-u_2/u_3)^{n_2} u_4^{n_3} u_3^{n_4}}{n_1! \, n_2! \, n_3! \, n_4!} 
\end{align}

\begin{align}
S_5=& u_3^{-2-\epsilon} \sum\limits_{n_3 > n_1 +n_4} \Gamma \left(n_1+n_2+1\right) \Gamma \left(-n_1+n_3-n_4\right) \Gamma \left(n_1+n_4+1\right) \Gamma \left(\epsilon +n_1+n_2-n_3+2\right)
\nonumber \\ & \times
  \Gamma\left(-\epsilon -n_1+n_3-1\right) \Gamma \left(-\epsilon -n_2+n_4-1\right)
 \frac{(-u_1 u_4 /u_3)^{n_1} (-u_2/u_3)^{n_2} (-u_3/u_4)^{n_3} (-u_4)^{n_4}}{n_1! \, n_2! \, n_3! \, n_4!} 
\end{align}

\begin{align}
S_6=& u_2^{-1-\epsilon} u_4^{-1} \sum\limits_{n_i=0}^{\infty}
\frac{\Gamma \left(n_2+n_3+1\right) \Gamma \left(n_1+n_4+1\right) \Gamma \left(\epsilon -n_3-n_4+1\right) \Gamma \left(-\epsilon +n_1+n_3+n_4\right)}{\Gamma(2+ n_1+n_2+n_3+n_4)}
\nonumber \\ & \times
   \Gamma \left(-\epsilon +n_2+n_3+n_4\right) \bigg(\psi\left(-\epsilon +n_2+n_3+n_4\right)-\psi\left(n_2+1\right) +\psi\left(n_2+n_3+1\right)
\nonumber \\ &   
   -\psi\left(n_1+n_2+n_3+n_4+2\right)+\log \left(u_3\right)-\log \left(u_4\right)\bigg)
 \frac{u_1^{n_1} (u_3/u_4)^{n_2} (u_2/u_4)^{n_3} u_2^{n_4}}{n_1! \, n_2! \, n_3! \, n_4!} 
\end{align}

\begin{align}
S_7=& u_2^{-1-\epsilon} u_3^{-1} \sum_{n_1 > n_2 +n_4} \Gamma \left(n_2+n_3+1\right) \Gamma \left(n_1-n_2-n_4\right) \Gamma \left(n_2+n_4+1\right) \Gamma \left(\epsilon +n_1-n_2-n_3-n_4+1\right)\nonumber \\ & \times
   \Gamma \left(-\epsilon -n_1+n_4-1\right) \Gamma \left(-\epsilon +n_2+n_3+n_4\right) \frac{(-u_1/u_2)^{n_1} (-u_2 u_4/u_3)^{n_2} (-u_2/u_3)^{n_3} (-u_2)^{n_4}}{n_1! \, n_2! \, n_3! \, n_4!} 
\end{align}

\begin{align}
S_8=& -u_2^{-1-\epsilon} u_3^{-1} \sum\limits_{n_1 \leq n_2 +n_4} 
\frac{\Gamma \left(n_2+n_3+1\right) \Gamma \left(n_2+n_4+1\right) \Gamma \left(\epsilon +n_1-n_2-n_3-n_4+1\right)}{\Gamma(1-n_1+n_2+n_4)} 
\nonumber \\ & \times
\Gamma \left(-\epsilon
   -n_1+n_4-1\right) \Gamma \left(-\epsilon +n_2+n_3+n_4\right) \bigg(-\psi\left(\epsilon +n_1-n_2-n_3-n_4+1\right)
   \nonumber \\ &
   +\psi\left(-\epsilon +n_2+n_3+n_4\right)-\psi\left(n_2+1\right)+\psi\left(n_2+n_3+1\right)+\psi\left(n_2+n_4+1\right)
    \nonumber \\ &
    -\psi\left(-n_1+n_2+n_4+1\right)
   +\log \left(u_2\right)-\log \left(u_3\right)+\log \left(u_4\right)\bigg)
 \frac{(u_1/u_2)^{n_1} (u_2 u_4/u_3)^{n_2} (- u_2/u_3)^{n_3} u_2^{n_4}}{n_1! \, n_2! \, n_3! \, n_4!} 
\end{align}

\begin{align}
S_9=& u_2^{-1-\epsilon} u_3^{-1} \sum_{\mathclap{\substack{n_3 > n_1 +n_4\\ 1+n_1+n_2+n_4 > n_3}}} \Gamma \left(-n_1+n_3-n_4\right) \Gamma \left(n_1+n_4+1\right) \Gamma \left(n_1+n_2-n_3+n_4+1\right) \Gamma \left(-\epsilon -n_1+n_3-1\right)
\nonumber \\ & \times
   \Gamma \left(\epsilon -n_2-n_4+1\right) \Gamma \left(-\epsilon +n_1+n_2+n_4\right) \frac{(-u_1 u_4/u_3)^{n_1} (-u_2 /u_3)^{n_2} (-u_3/u_4)^{n_3} (-u_2 u_4/u_3)^{n_4}}{n_1! \, n_2! \, n_3! \, n_4!} 
\end{align}

\begin{align}
S_{10}=& u_1^{-1-\epsilon} u_3^{-1} \sum_{n_2>n_1+n_3} \Gamma \left(-n_1+n_2-n_3\right) \Gamma \left(n_1+n_3+1\right) \Gamma \left(n_2+n_4+1\right) \Gamma \left(-\epsilon -n_2+n_3-1\right)
\nonumber \\ & \times
   \Gamma
   \left(\epsilon -n_3-n_4+1\right) \Gamma \left(-\epsilon +n_1+n_3+n_4\right) \frac{(-u_2/u_3)^{n_1} (-u_4)^{n_2} (-u_1/u_3)^{n_3} (-u_1)^{n_4}}{n_1! \, n_2! \, n_3! \, n_4!} 
\end{align}

\begin{align}
S_{11}=& -u_1^{-1-\epsilon} u_3^{-1} \sum\limits_{n_2 \leq n_1 +n_3} 
\frac{\Gamma \left(n_1+n_3+1\right) \Gamma \left(n_2+n_4+1\right) \Gamma \left(-\epsilon -n_2+n_3-1\right)}{\Gamma(1+n_1-n_2+n_3)} 
\nonumber \\ & \times
\Gamma \left(\epsilon -n_3-n_4+1\right)
   \Gamma \left(-\epsilon +n_1+n_3+n_4\right) \bigg(-\psi\left(-\epsilon +n_1+n_3+n_4\right)+\psi\left(n_1+1\right)
   \nonumber \\ &-\psi\left(n_1+n_3+1\right)+\psi
   \left(n_1-n_2+n_3+1\right)-\log \left(u_2\right)+\log \left(u_3\right)\bigg)
 \frac{(u_2/u_3)^{n_1} u_4^{n_2} (u_1/u_3)^{n_3} (-u_1)^{n_4}}{n_1! \, n_2! \, n_3! \, n_4!} 
\end{align}

\begin{align}
S_{12}=& (u_1 u_4)^{-1-\epsilon} u_3^{-1} \sum_{n_i=0}^{\infty}\Gamma \left(n_1+n_2+1\right) \Gamma \left(\epsilon -n_2-n_3+1\right) \Gamma \left(-\epsilon +n_1+n_2+n_3\right) \Gamma \left(\epsilon
   -n_2-n_4+1\right)
\nonumber \\ & \times
 \Gamma \left(-\epsilon -n_1+n_4-1\right) \Gamma \left(-\epsilon +n_2+n_3+n_4\right) \frac{(-u_2/u_3)^{n_1} (-u_1 u_4/u_3)^{n_2} (-u_1)^{n_3} (-u_4)^{n_4}}{n_1! \, n_2! \, n_3! \, n_4!} 
\end{align}

\begin{align}
S_{13}=& u_1^{-1-\epsilon} u_3^{-1} \sum_{n_2>n_1+n_3} \Gamma \left(-n_1+n_2-n_3\right) \Gamma \left(n_1+n_3+1\right) \Gamma \left(n_1+n_4+1\right) \Gamma \left(-\epsilon -n_1+n_2-1\right)
\nonumber \\ & \times
   \Gamma
   \left(\epsilon -n_2-n_4+1\right) \Gamma \left(-\epsilon +n_1+n_3+n_4\right) \frac{(-u_2 u_4/u_3)^{n_1} (-u_1/u_2)^{n_2} (-u_2/u_3)^{n_3} (-u_1)^{n_4}}{n_1! \, n_2! \, n_3! \, n_4!} 
\end{align}

\begin{align}
S_{14}=& (u_1 u_2 u_4)^{-1-\epsilon} u_3^{\epsilon} \sum_{n_i=0}^{\infty}\Gamma \left(\epsilon -n_1-n_3+1\right) \Gamma \left(\epsilon -n_1-n_4+1\right) \Gamma \left(\epsilon -n_2-n_4+1\right) 
\nonumber \\ & \times
 \Gamma \left(-\epsilon
   +n_1+n_2+n_4\right)\Gamma \left(-\epsilon +n_1+n_3+n_4\right) \Gamma \left(-2 \epsilon +n_1+n_2+n_3+n_4-1\right) 
   \nonumber \\ & \times \frac{(-u_1 u_4/u_3)^{n_1} (-u_2/u_3)^{n_2} (-u_1)^{n_3} (-u_2 u_4/u_3)^{n_4}}{n_1! \, n_2! \, n_3! \, n_4!} 
\end{align}

\end{document}